# Magnetometry using fluorescence of sodium vapor

TINGWEI FAN,[1,2] LEI ZHANG,[1] XUEZONG YANG,[1,2] SHUZHEN CUI,[1] TIANHUA ZHOU,[1,3] AND YAN FENG[1,*]

[1] *Shanghai Key Laboratory of Solid State Laser and Application, and Shanghai Institute of Optics and Fine Mechanics, Chinese Academy of Sciences, Shanghai 201800, China.*

[2]*University of the Chinese Academy of Sciences, Beijing 100049, China.*

[3]*siomzth@siom.ac.cn*

**Corresponding author: feng@siom.ac.cn*

**Magnetic resonance of sodium fluorescence is studied with varying laser intensity, duty cycle, and field strength. A magnetometer based on sodium vapor cell filled with He buffer gas is demonstrated, which uses a single amplitude-modulated laser beam. With a 589 nm laser tuned at $D_1$ or $D_2$ line, the magnetic field is inferred from the variation of fluorescence. A magnetic field sensitivity of 150 pT/$\sqrt{Hz}$ is achieved at $D_1$ line. The work is a step towards sensitive remote magnetometry with mesospheric sodium.**

In the past two decades, atomic magnetometers research has made significant advances. The sensitivity of atomic magnetometers rivals [1, 2] and even surpasses [3, 4] that of superconducting quantum interference device (SQUID)-based magnetometers. Magnetometer based on vapor of alkali atoms (Rb, Cs or K) has been a research focus [5-7]. Sodium atoms were initially used in various spectral studies [8, 9]. However, magnetometer using sodium atom had not been investigated because of its relatively low vapor pressure and lackness of suitable pump laser source.

In 2011, Higbie *et. al.* proposed a method for remotely measuring the geomagnetic field by interrogating sodium atoms in the mesosphere with a pulsed laser at the sodium resonance wavelength [10]. Measurement of magnetic field on the 100 km length scale is significant for many geophysical applications including mapping of crustal magnetism and ocean circulation measurements. Kane *et. al.* demonstrated an initial measurement based on mesospheric sodium in 2016. A measurement sensitivity of 162 nT/$\sqrt{Hz}$ was reported [11]. Considering the complexity and cost, experimental study in the lab is beneficial before the on-sky test. Under laboratory condition, it is easier to make subtle investigation with higher sensitivity. Indeed it is difficult to mimic the physical condition of the mesospheric sodium layer in the lab. But the same quantum law governs the light and sodium atom interaction either in 90 km high altitude or in a vapor cell on the ground. Laboratory study of sodium magnetometry is a necessary step towards the realization of high sensitivity remote magnetometry with mesospheric sodium.

In this paper, we present a study on the magnetic resonance of sodium fluorescence and a demonstration of sodium magnetometer. Considering the feasibility in extending the scheme to mesospheric sodium later, fluorescence signal from a sodium vapor cell is used for the measurement. Buffer gas is filled to increase the spin polarization lifetime by slowing the diffusion of atomic sodium to the cell walls. The magnetometer, which uses a single amplitude-modulated laser beam, is a type of Bell–Bloom magnetometer[12]. By modulating the laser amplitude around the Larmor frequency, a magnetic resonance happens because of the coherence reinforced by synchronizing optical pumping with the spin precession. A magnetic field sensitivity of 150 pT/$\sqrt{Hz}$ is achieved at Na $D_1$ line.

The schematic diagram of experimental setup is shown in Fig. 1. The light source is a frequency-doubled diode-seeded Raman fiber amplifier, which provides up to 1 W continuous wave 589 nm laser with a measured linewidth of ~ 5 MHz (similar to the one reported in [13]). A wavelength meter is used to monitor and control the center wavelength of the laser in real time, with an absolute accuracy of 200 MHz. After collimation, the linearly polarized laser is amplitude modulated by an acousto-optic modulator (AOM). The first order diffraction is chosen to work with, because its extinction ratio is higher than the zero-order. A quarter-wave plate changes the laser from linear to circular polarization. The laser beam is expanded to 3.7 mm diameter, and then illuminates the cylindrical sodium vapor cell whose radius and length are 1.5 cm and 5 cm, respectively. The sodium vapor cell containing enriched Na with 100 Torr He buffer gas, which was heated to 60°C by channeling warm air into the cell oven. The cell oven is inside a four layer magnetic shield that provides nearly isotropic shielding of external magnetic fields. An adjustable magnetic field is generated along the axial direction of the cylindrical magnetic shield with a magnetic coil. In the experiments, the laser beam direction is perpendicular to the magnetic field at the vapor cell.

The fluorescence of the sodium vapor is detected through an axial hole of the magnetic shield. In the magnetic resonance study, a photon counting head (Hamamatsu H10682-210) is used with a photon counter (Hamamatsu C8855-01). A 589 nm filter with a full width at half maximum (FWHM) linewidth of 1nm (Alluxa 589.45-1 OD6 ULTRA Bandpass Filter) is put before the detector to reduce the background light. Alternatively, a photomultiplier tube (Hamamatsu R9880U-210) and lock-in amplifier (Stanford Research Systems SR830) are used to measure the magnetic field and analyze the measurement sensitivity. A modulated pulse signal is applied to drive the AOM, and the sync output of the signal generator acts as the reference signal of the lock-in amplifier. The lock-in amplifier demodulates the experimental signal at the reference frequency.

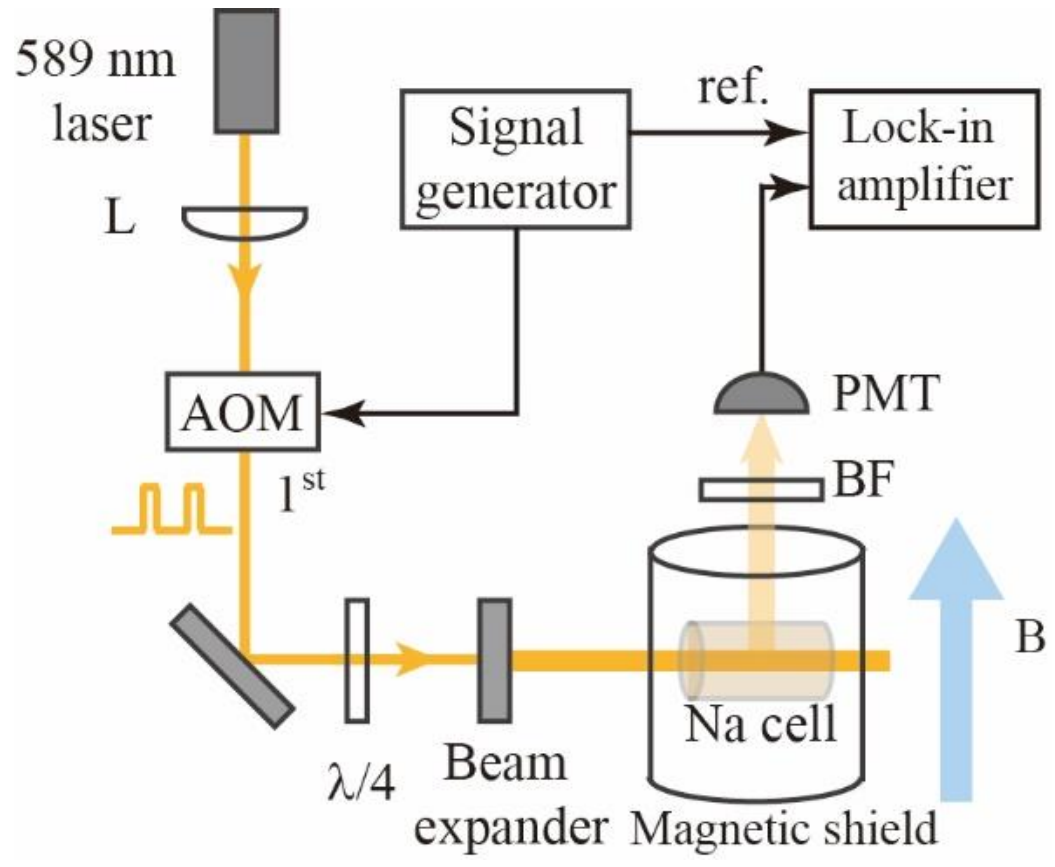

Fig. 1. Schematic of the experimental arrangement for sodium magnetometer with amplitude modulation. L, lens; λ/4, quarter-wave plate; AOM, acousto-optic modulator; PMT, photomultiplier tube; BF, bandpass filter. The temperature-controlled Na vapor cell is immersed in magnetic field that is along the axis of the magnetic shield. The signal generator generates a modulated pulse signal to drive the AOM. The lock-in amplifier demodulating the fluorescence signal at the synchronous modulation frequency.

The sodium fluorescence is measured in the direction perpendicular to the laser beam. Measuring at the backward direction is more consistent with the expected setup of remote magnetometry with mesospheric sodium. However, trouble with backscattered stray light was encountered in the experiments. The perpendicular direction was chosen instead for less stray light. Special attention was paid to decrease the background illumination of the photon detector from the scattered radiation of the laser.

Because of the dark magnetic resonance and correspondingly reduced photon shot noise, $D_1$ line is expected to allow better sensitivity[10]. The laser was therefore tuned to the sodium $D_1$ line (589.7558 nm). The sodium atom density is low at room temperature ($7.89\times10^5$ atoms/cm$^3$ at 25°C[14]). At the temperature of 60°C, the sodium density is approximately $6.65\times10^7$ atoms/cm$^3$. The spin relaxation time is calculated to be 48.2 ms (considering the 60 °C temperature, the 100 Tor He buffer gas, and the cell dimension [15, 16]). The FWHM width of the sodium $D_1$ line is measured to be 2.8 GHz at the experimental condition, as seen in section 1 of Supplement 1. The spectral line is collision broadened strongly that the hyperfine structure is blurred. In the experiments, best results are obtained with the laser tuned at 589.7558 nm which is the fine structure wavelength of the $D_1$ line.

Rectangular pulse signals are applied to the AOM which produces an amplitude modulated laser. The repetition frequency is scanned across the Larmor frequency. Typical waveforms are shown in section 2 of the Supplement 1. The fluorescence photons are counted synchronously with the pulse repetition frequency scan. Magnetic resonance lines were recorded for different duty cycle and peak intensity of the laser, and varying magnetic field. A typical resonance line is shown in Fig. 2. The lineshape is close to Lorentzian. For the magnetic resonance shown in Fig. 2, the Larmor frequency $\nu_L$ and the FWHM of magnetic resonance $\Delta f$ are 297.17 kHz and 580 Hz, respectively. The magnetic field $B$ is $\nu_L/g$ = 0.4246 G, where $g = 6.99812$ Hz/nT is the gyromagnetic ratio of atomic sodium. The shot-noise-limited (SN-limited) sensitivity is calculated to be 147 pT/$\sqrt{Hz}$ with the following formula [11],

$$\delta B_{SN} = 4\sqrt{3}\Delta f/9g \times \left(\sqrt{S_{Dip} + S_{Back}}\right)/\left(\left|S_{Dip} - S_{Back}\right|\right) \quad (1)$$

Where $S_{Dip}$ and $S_{Back}$ are the photon detection rates at resonance and background, respectively. The formula is valid for the case that the resonance shape is Lorentzian. It is used to quickly assess the magnetic resonance signal and find optimum condition for sensitive magnetic field measurement.

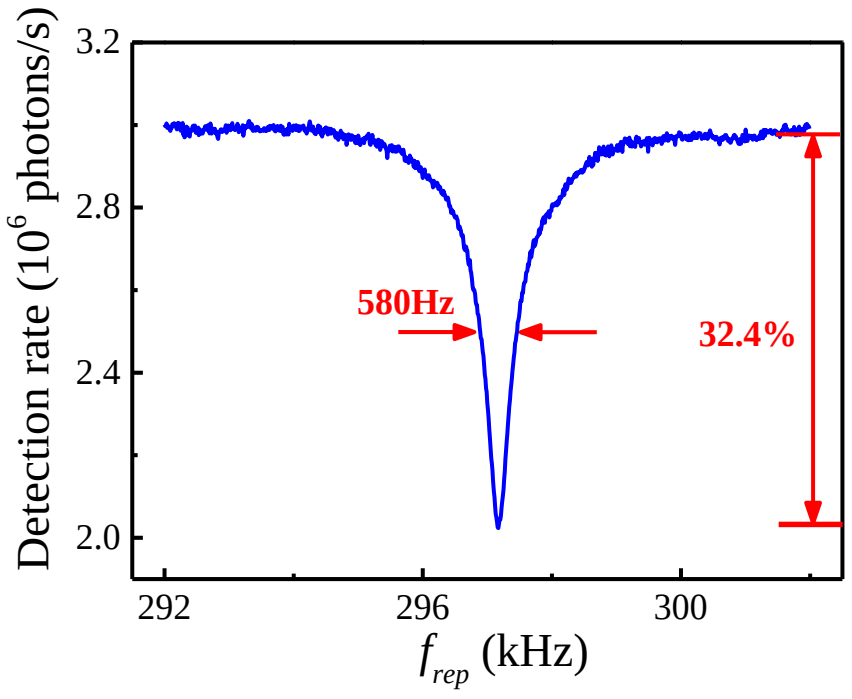

Fig. 2. Typical magnetic resonance signal of Na $D_1$ line, where the peak laser intensity is 65 W/m$^2$, pulse duty cycle is 17.5%, and cell temperature is 60°C.

Fig. 3 shows plots of magnetic resonance width, resonance height, and calculated SN-limited sensitivity as functions of laser peak intensity and pulse duty cycle. In these experiments, the magnetic field is fixed at 0.4246 G. When varying the laser peak intensity as in Fig. 3 (a), the pulse duty cycle is fixed at 20 %. It is found that the magnetic resonance width increases linearly with the increase of the peak intensity. The relative resonance height is saturated at high peak intensity. The highest SN-limited sensitivity is calculated at a laser intensity of 65 W/m$^2$. Then in the pulse duty cycle optimization in Fig. 3 (b), the laser intensity is fixed at 65 W/m$^2$. The resonance width increases with the duty cycle. The resonance height depends on the duty cycle strongly and has an optimum around 17.5%. As a result, a close to optimum condition is determined at a duty cycle of 17.5% and peak intensity of 65 W/m$^2$. The respective resonant fluorescence signal is shown in Fig.2 and discussed in previous paragraph.

The magnetic resonance not only happens at the Larmor frequency but also at the subharmonics of the Larmor frequency. While the resonance at the Larmor frequency corresponds to the physical case that the sodium atoms are excited at every Larmor period, the resonances at subharmonics correspond to excitation at every multiple Larmor periods. Figure 4 (a) plots the multiple resonance signal when the laser duty cycle is 20 %. Second and third subharmonics are clearly observed. More subharmonics can be observed with smaller duty cycle. Figure 4 (b) plots the case when the magnetic field is 515.7 nT, much smaller than that in Fig. 4 (a). It is found that the Larmor resonance and subharmonic resonance start to overlap with each other. At even smaller magnetic field, the resonance width cannot be read accurately so that simple estimation of the SN-limited sensitivity with Equ. 1 is not feasible.

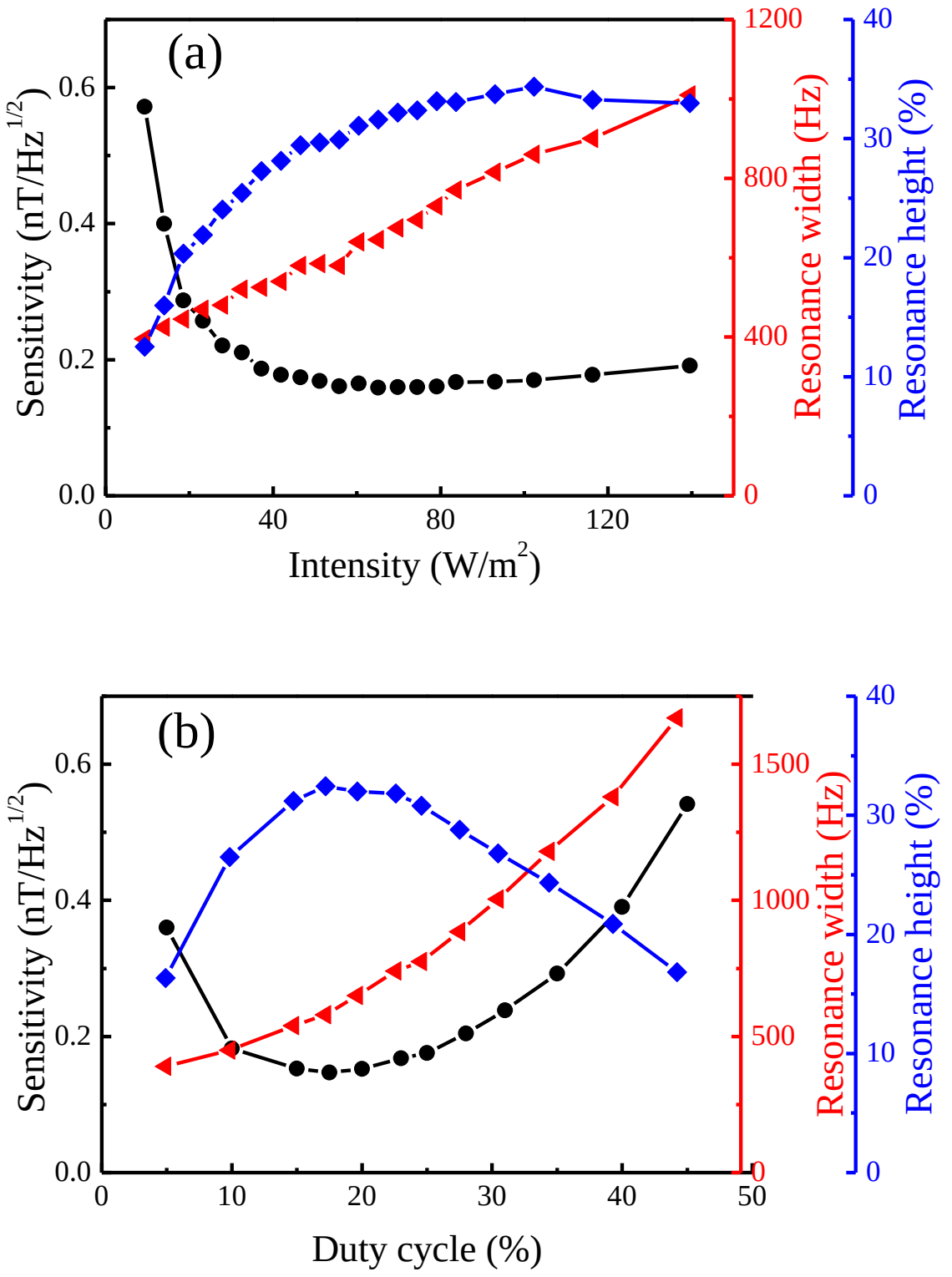

Fig. 3. Plots of magnetic resonance width, relative resonance height, and SN-limited sensitivity as functions of the laser peak intensity and pulse duty cycle. Symbols are the experimental data for the magnetic resonances of the $D_1$ sodium lines at different parameters. (a) Dependence on peak intensity at a pulse duty cycle of 20%. (b) Dependence on duty cycle at a fixed laser intensity of 65 W/m².

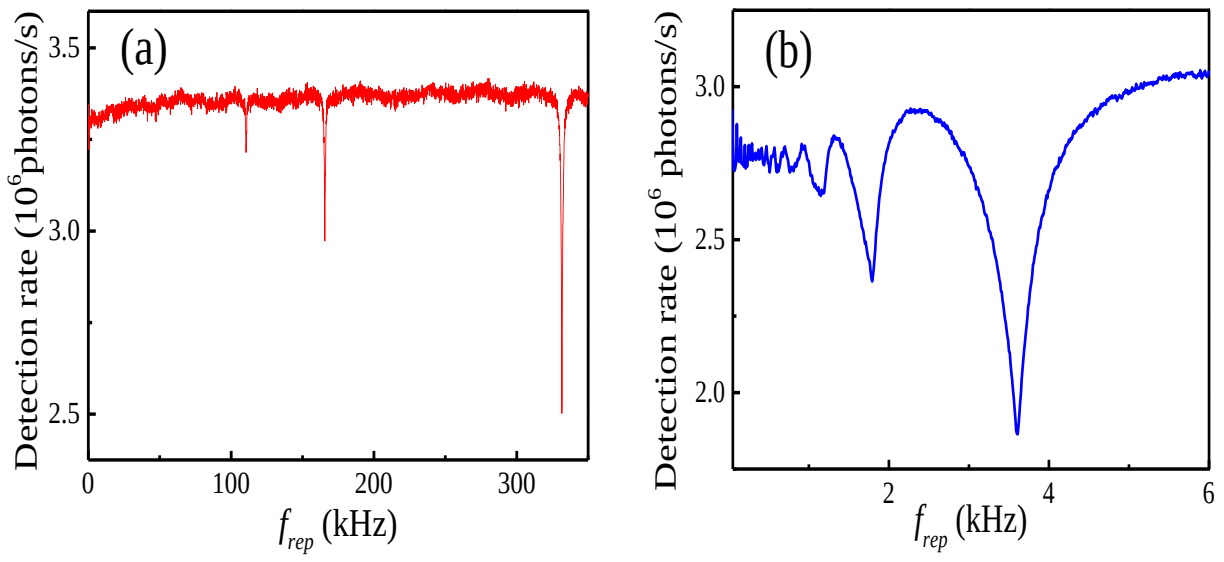

Fig. 4. Multiple magnetic resonance signal at a duty cycle of 20% and peak laser intensity of 65 W/m². (a) magnetic field of 47371.4 nT. (b) magnetic field of 515.7 nT.

Fig. 5 shows plots of magnetic resonance width, relative resonance height, and calculated SN-limited sensitivity as functions of magnetic field, where the laser duty cycle is 20% and the peak intensity is 65 W/m². With decreasing magnetic field, the resonance width decreases, the resonance height increases, and so the SN-limited sensitivity improves. At lower magnetic field, the resonance width and relative height almost keep constant. The SN-limited sensitivity is about 122 $\mathrm{pT}/\sqrt{Hz}$. For magnetic field less than 500 nT, the resonance signal can still be clearly observed, although the Larmor resonance and subharmonic resonance overlap more. Resonance signal for magnetic field as low as <20 nT can be detected (see section 3 of Supplement 1). Therefore, the setup can be used to detect not only geomagnetic field but also small field. The minimum detectable field is determined by the spin relaxation time of sodium atoms. The spin relaxation time is estimated to be 48.2 ms in our experimental condition. To drive a coherence, the relaxation rate should at least a few times lower than the spin-precession frequency, so a minimum detectable magnetic field of about 20 nT is reasonable.

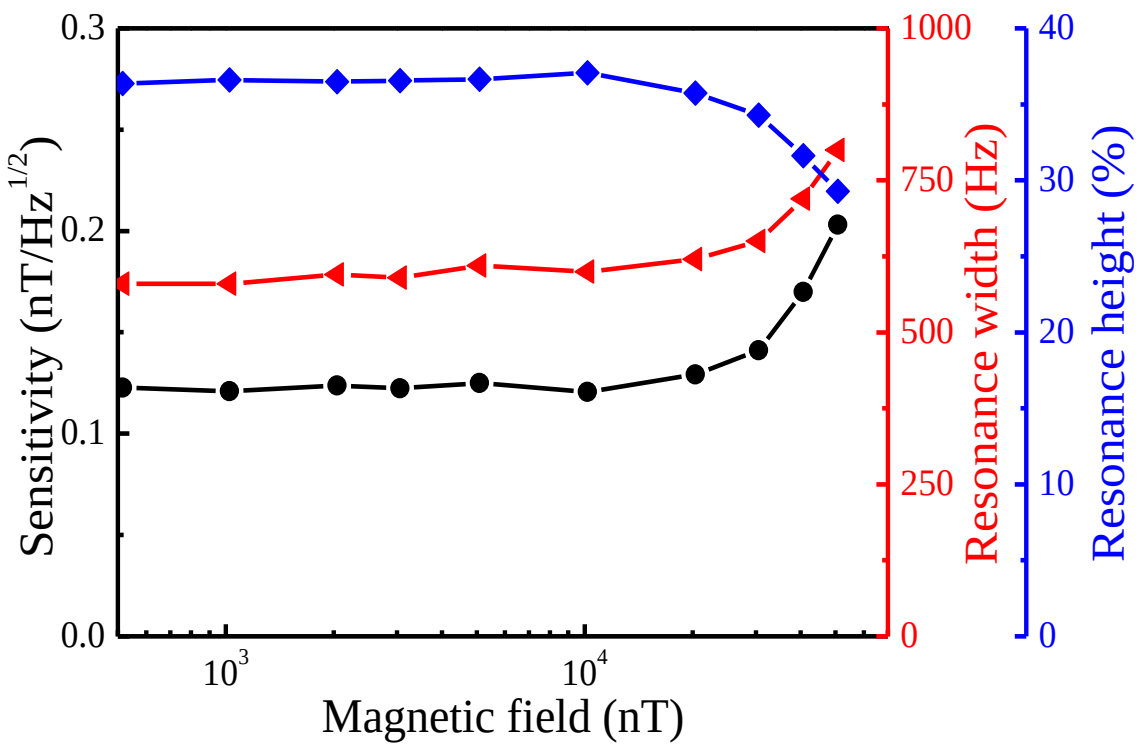

Fig. 5. Magnetic field dependence of magnetic resonance width, relative resonance height, and calculated SN-limited sensitivity, where the laser duty cycle is 20% and the peak intensity is 65 W/m².

In order to make a sensitive measurement of the magnetic field, we utilize a phase sensitive technique[17]. The detector is switched to a photomultiplier tube with analog output. The laser pulse repetition frequency $f_{rep}$ is modulated around a central frequency $f_c$ : $f_{rep} = f_c + f_{dev}cos(2\pi f_{mod}t)$, where $f_{mod}$ is the modulation frequency and $f_{dev}$ is the amplitude of modulation. The detected signal is demodulated with a lock-in amplifier referenced to the modulation frequency $f_{mod}$. The output of the lock-in amplifier is linear $S_{LI} \approx \alpha(f_c - f_{res})$ as a function of $f_c - f_{res}$ when $|f_c - f_{res}| < \Delta f/2$ , where $f_{res}$ is resonance frequency. The slope $\alpha \propto l/\Delta f$ is extracted from the trace in Fig. 6(a), which shows dispersive shaped resonance from the X output of the lock-in amplifier, where $l$ is the contrast of magnetic resonance. The slope $\alpha$ depends on $f_{mod}$. The modulation frequency should be slow enough to allow adequate optical pumping. Meanwhile the modulation frequency should be fast enough because of the bandwidth requirement and the increasing intensity noise at low-frequency of the laser. We tested multiple modulation frequencies and 500 Hz was chosen.

The lock-in signal is linear for small changes $\Delta B$, when the central repetition frequency is set to a certain value around $f_{res}$. Spectral dependence of the noise contributions are determined from a radio frequency analysis of the X output of the lock-in amplifier (Stanford Research SR785 Dynamic Signal Analyzer). The root-mean-square (rms) magnetic field fluctuations are shown in Fig. 6(b). For frequencies in the range of 1–100 Hz, a noise floor of 150 $\mathrm{pT}/\sqrt{Hz}$ is reached. For frequencies below 1 Hz additional noise is present. Above 100 Hz the noise decreases due to the finite time constant of the lock-in amplifier (here 1ms), which reduces the bandwidth to ~ 300 Hz. Several noise peaks are also present, which are at the line frequency and its harmonics.

There are several measures which can be used to potentially reduce the noise floor or sensitivity. The beam spot size is only 3.7 mm right now. With bigger beam spot, the number of the interacting sodium atoms will increase. The optical setup for fluorescence detection can be improved for collecting more photons. Moreover, there are technical noise sources related to the power and wavelength stability of the pump laser. We are going to replace the laser with one of better power stability. The wavelength will be locked with Doppler-free saturated absorption spectroscopy, which will improve the wavelength stability greatly comparing to current wavemeter based method.

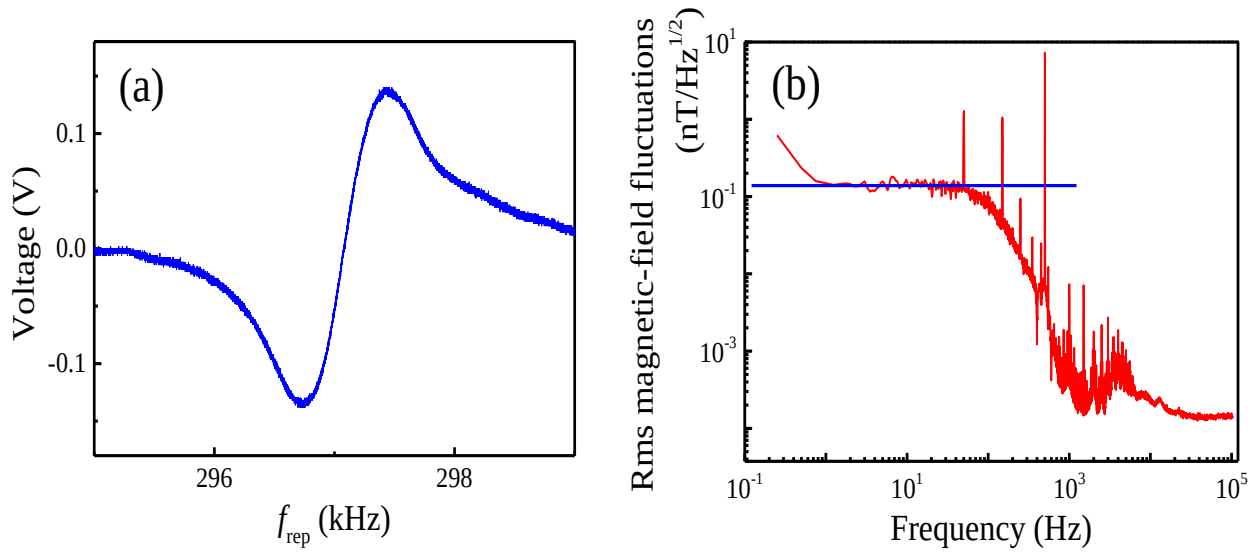

Fig. 6. Measurement of magnetic field with PMT and lock-in amplifier with a modulation frequency of 500 Hz. (a) Lock-in signal as a function of the central repetition frequency $f_c$ which is scanned over a magnetic resonance. (b) Magnetic field noise spectrum.

We have successfully demonstrated magnetometry with sodium vapor by detecting the fluorescence. Nevertheless, there is an observation we could not understand yet. There is a small dip on the high frequency side of the magnetic resonance. This feature becomes obvious at lower laser power, where the resonance width decreases. When the laser intensity is low enough that the resonance width is less than its spacing from the dip, the dip can be resolved. A typical resonance signal with the side feature is shown in Fig. 7 (a). The frequency spacing between the dip and center frequency of the magnetic resonance is measured with respect to the magnetic field. A linear dependence is found as shown in Fig. 7 (b).

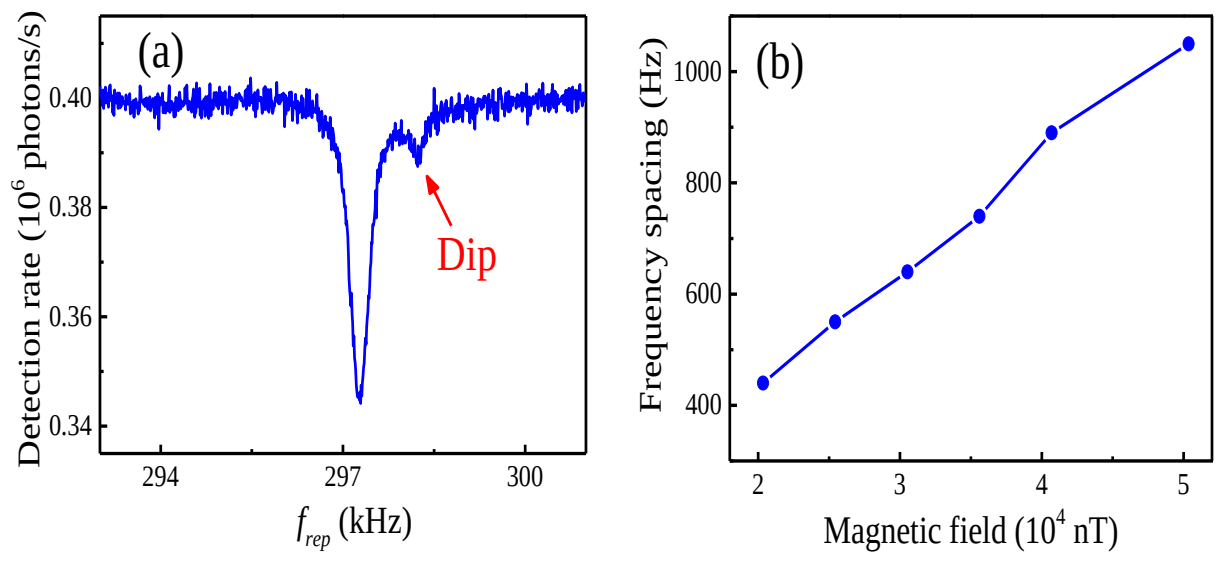

Fig. 7. (a) Typical resonance line at small light intensity for Na $D_1$ line. (b) Frequency shift of the small dip relative to the resonance frequency as a function of magnetic field.

Magnetic resonance of sodium fluorescence pumped at $D_2$ line (589.1583 nm) was investigated as well. While a peak signal was expected according to previous simulation [10], a valley was observed (see section 4 of Supplement 1). The center frequency of the valley signal varies with the magnetic field, which confirms that it is the magnetic resonance signal. We believe the observation is related to the direction of the fluorescence detection. At good optical pumping condition, as is at the magnetic resonance, a majority of the sodium fluorescence is circularly polarized. The pattern of circular polarization emission is peanut-shaped (see section 4 of Supplement 1). Therefore, the emission to the perpendicular direction is actually reduced as compared to off-resonance condition, which results in a valley at the resonance. At off-resonance condition, the precession of sodium atoms in magnetic field destroys the effect of optical pumping. The circular polarization fluorescence loses majority. A direct confirmation would be a measurement at the backward direction, which however cannot be achieved in our current setup.

In summary, we have investigated the magnetic resonance of sodium fluorescence and demonstrated a magnetometer based on sodium vapor cell with He buffer gas. The setup can be used to measure magnetic field ranging from ~ 20 nT to the geomagnetic field and higher. The magnetometer, which uses the technique of single amplitude-modulated laser beam, is a type of Bell–Bloom magnetometer. A 589 nm laser illuminates the atoms, and the magnetic field is inferred from fluorescence collected at a photon detector. Magnetic resonance is detected at both $D_1$ and $D_2$ lines. A magnetic field sensitivity of 150 pT/$\sqrt{Hz}$ is demonstrated at $D_1$ line.

Considerable improvement can be expected, by improving the power and wavelength stability of the laser, and increasing the involving sodium atoms with a bigger beam spot and optimized fluorescence detecting optics. More attention should be devoted to decreasing the background illumination of the detector from the scattered radiation of the laser. Although the buffer gas slows down the diffusion of alkali atoms to the cell wall, collisions with buffer gas atoms broaden the spectral width and limit magnetometer performance, especially in high-pressure cells. If possible, the cell filled with buffer gas should be replaced by coated cells that allow for easier realization of high magnetic field sensitivity. Future work will be magnetometry with mesospheric sodium. But before that, a fine tuning of the experimental condition to mimic the remote magnetometry would be also useful.

See Supplement 1 for supporting content.

# Supplement 1

### 1. Spectrum of the sodium $D_1$ transition

By counting the 589 nm photons with the laser scanned across the sodium line, spectra are obtained which contain information of sodium atom absorption and background scattered light. A spectrum for $D_1$ line taken at a temperature of 60°C and a peak intensity of 65 W/m$^2$ is shown in Fig.S1. The spectrum is centered at 589.7558 nm, and the linewidth is ~ 2.8 GHz which is collision broadened. The background scattered light is about 21% of the total signal.

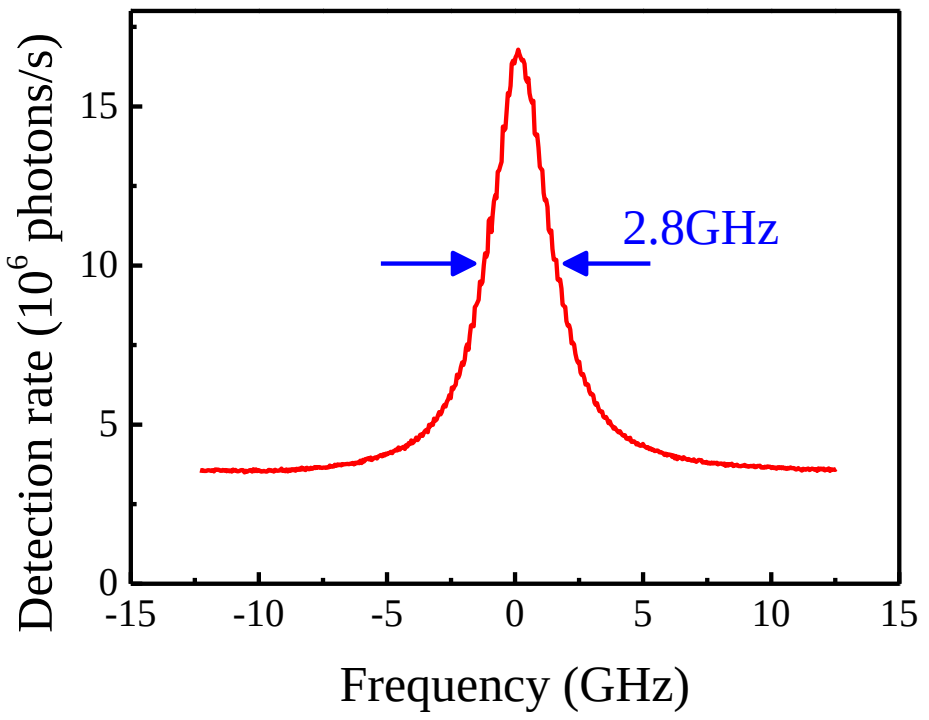

Fig. S1. A spectrum for $D_1$ line taken at a temperature of 60 °C and a peak intensity of 65 W/m$^2$.

### 2. Amplitude modulated output of the laser

Rectangular pulse signals are applied to the AOM, which produces an amplitude modulated laser. Waveforms for repetition frequency of 250kHz and 300kHz and duty cycle of 20% are shown in Fig. S2. Fluctuations of laser peak power are obvious, which is one of the factors limiting the sensitivity.

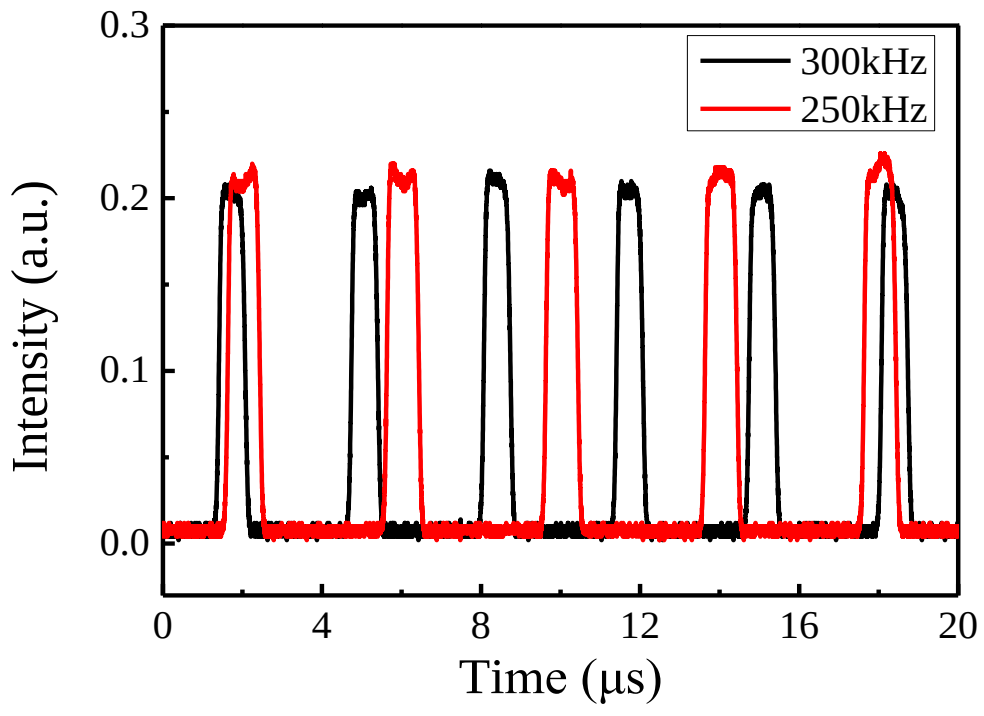

Fig. S2. Pulse trains of the amplitude modulated laser at a duty cycle of 20%.

### 3. Resonance signal for magnetic field as low as tens of nanotesla

Fig. S3 plot the resonance signals for magnetic field as low as tens of nanotesla when the laser duty cycle is 20 % and peak intensity is 65 W/m$^2$. Fig. S3 (a) plots the case for a magnetic field of 10 nT. Because the Larmor frequency (about 70 Hz) becomes close to the photon counting frequency (here 5 Hz), fluctuation of detection rate is obvious. The magnetic resonant signal is visible but shallow. Increasing the magnetic field slightly, the magnetic resonant signal becomes evident. Figure S3 (b) shows the case for 17.8 nT. Therefore, the setup can be used to detect magnetic field as low as <20 nT.

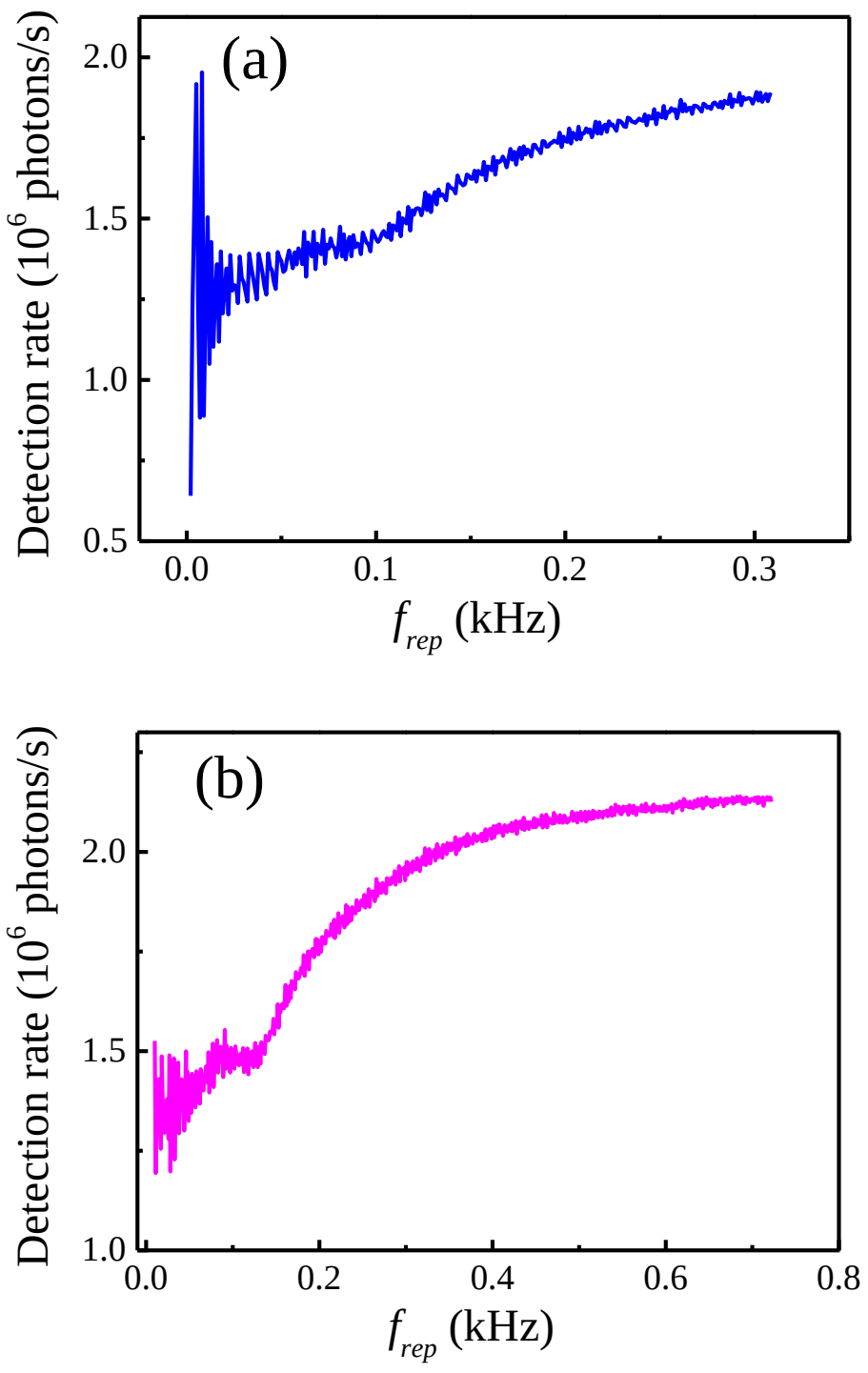

Fig. S3. Magnetic resonance signal at a duty cycle of 20% and peak laser intensity of 65 W/m$^2$. (a) magnetic field of 10 nT. (b) magnetic field of 17.8 nT.

### 4. Magnetic resonance signal for sodium $D_2$ line

When the laser is tuned to the sodium $D_2$ line (589.1583 nm) with the same set-up, the resonant signal we got is a valley as well. Typical resonance lines are shown in Fig. S4. The resonant center frequency varies with the values of the magnetic field, which confirm that it is the magnetic resonance signal. This observation is not expected before the experiments, because a peak resonance is calculated both in literature [1] and our own simulation. We believe the different observation is due to the different direction of detection. Fig. S5 shows the emission patterns for atoms spontaneously emitting linearly and circularly polarized photons, after excitation by a vertically oriented laser beam [2]. The circularly polarized light is more likely to be directed backward and has less probability to the perpendicular direction, while the linearly polarized emission is opposite. At good optical pumping condition, as is at the magnetic resonance, a majority of the sodium fluorescence is circularly polarized. This is the reason that a valley is observed in spite of the higher absorption at resonance.

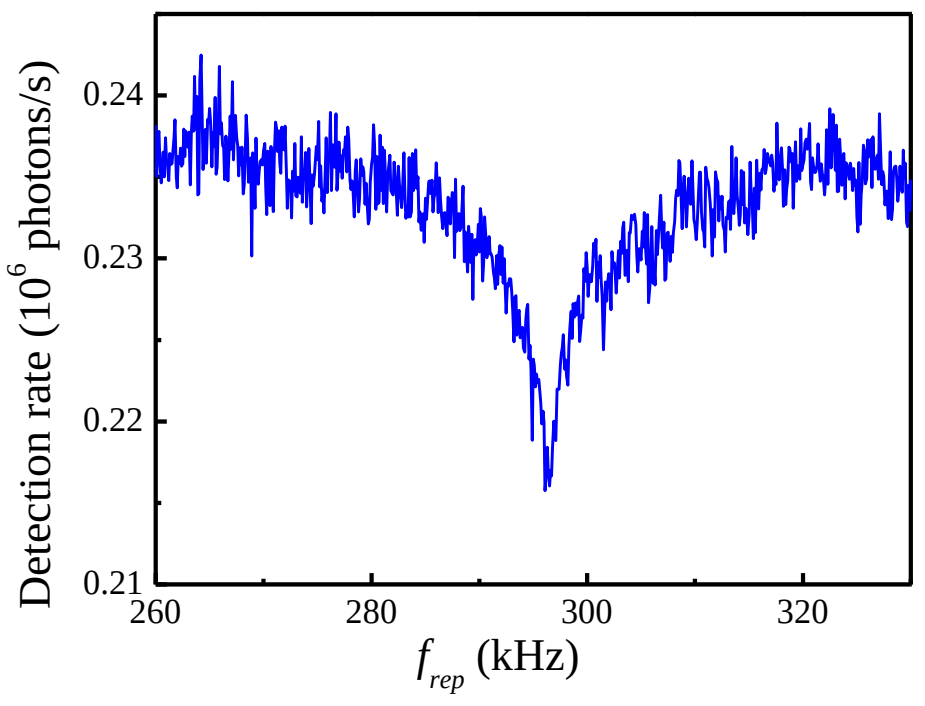

Fig. S4. Experimental magnetic resonance signal at sodium $D_2$ line.

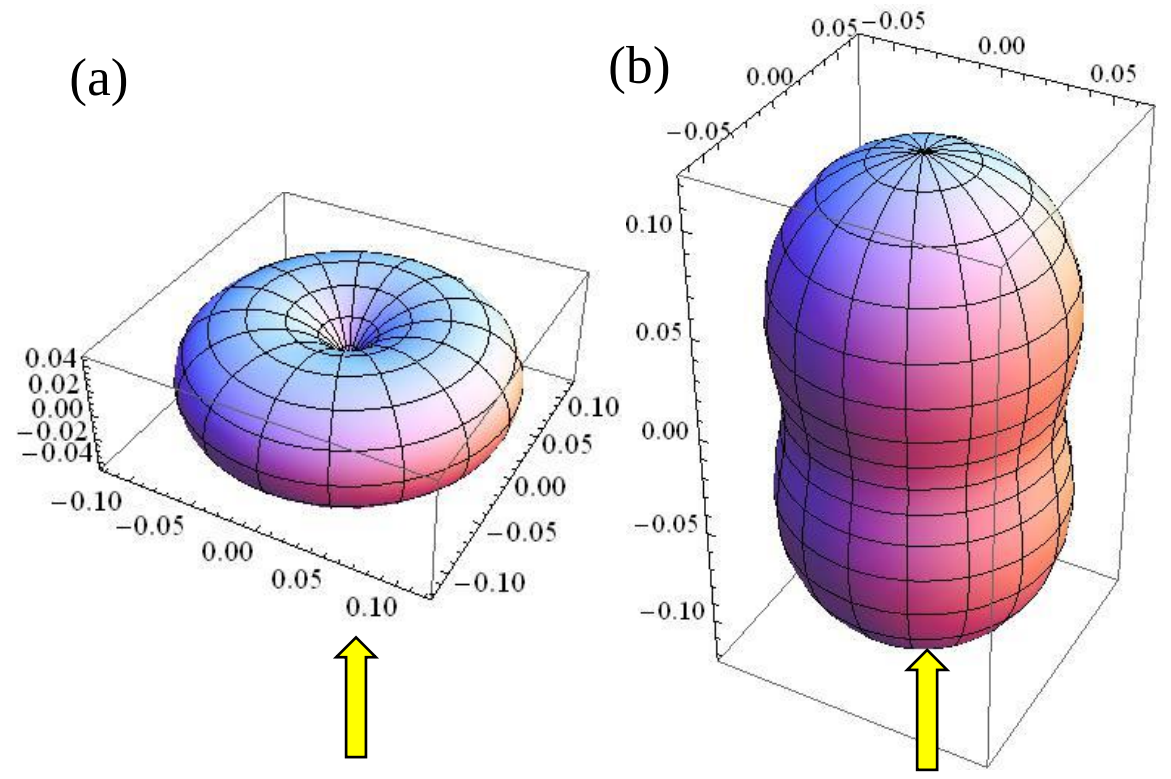

Fig. S5. Emission patterns for sodium $D_2$ transitions, in which the emitted photon is linearly polarized (a) or circular polarized (b). The yellow arrows indicate the direction of laser propagation.